\documentclass[a4paper,12pt]{article}
\usepackage{amsmath,amsthm,amsfonts,amssymb,latexsym,xy}
\usepackage[cp1251]{inputenc}  % win
\usepackage[english]{babel}
\usepackage{amsfonts,graphicx}
\usepackage{array,longtable,curves}
\usepackage{pstricks-add,xcolor}

 \numberwithin{equation}{section}

 \allowdisplaybreaks

\newtheorem{Pa}{Paper}[section]
\newtheorem{Tm}[Pa]{{\bf Theorem}}
\newtheorem{La}[Pa]{{\bf Lemma}}
\newtheorem{Cy}[Pa]{{\bf Corollary}}
\newtheorem{Rk}[Pa]{{\bf Remark}}

\newtheorem{Ee}[Pa]{{\bf Example}}

\begin{document}

%\noindent {\bfseries \Large \textsc{Arbitrage free markets geometry}}
\title{Banach geometry of arbitrage free markets}

%\medskip
\bigskip

\author{A.V. Lebedev\thanks{
\emph{e-mail:} \texttt{lebedev@bsu.by} (the corresponding author)}\\
Institute of Mathematics,  University  of Bialystok,\\
ul. Akademicka 2, PL-15-267  Bialystok, Poland\\
Department of Mechanics and Mathematics,
Belarus State University, \\
pr. Nezavisimosti, 4, 220050, Minsk, Belarus\\
\and
P. P.  Zabreiko\thanks{\emph{e-mail:} \texttt{zabreiko@mail.ru}}\\
Department of Mechanics and Mathematics,
Belarus State University, \\
pr. Nezavisimosti, 4, 220050, Minsk, Belarus}

\maketitle

%\footnote[1]{
%\emph{e-mail:} \texttt{lebedev@bsu.by} (the corresponding author)\\

%\ead{lebedev@bsu.by}

\begin{abstract}
 The article presents a description of  geometry of Banach  structures forming mathematical base of markets arbitrage absence type   phenomena.  In this connection the role of reflexive subspaces (replacing classically considered finite-dimensional subspaces) and plasterable cones is uncovered.
\end{abstract}

\medbreak


\medbreak

\tableofcontents

\section*{Introduction}\label{s.1}

The article presents analysis of geometry of Banach structures that can be interpreted as certain arbitrage free markets type phenomena.

In the theory  of Mathematical Finance the principal role in this field  is played by  the so-called 'Fundamental Theorem of asset Pricing' (in fact there is a series of results under this name related to situations in question). The Fundamental Theorem of asset Pricing  links arbitrage free markets (i.e markets that do not admit riskless claims yielding profit with strictly positive probability; the accurate  definition will be given below in Section~\ref{s.2}) with existence of martingales generated by measures that are equivalent to the initial one.
One should mention quite a number of researchers who contributed to the theme. Among them are F.~Black, M.~Scholes, R.~Merton, J.~Harrison, S.~Pliska,  S.~Ross, D.M.~Kreps, R.~Dalang, A.~Morton, W.~Willinger, D.~Kramkov, J.~Jacod, A.N.~Shiryaev,  D.R.~Rokhlin, F.~Delbaen, W.~Schachermayer, Yu. Kabanov  and many others.
We cannot give a full account of  sources and names related to the subject  and refer, for example,  to  \cite{Del-Schach},   \cite{Schach}, and \cite {Pasc} and the sources quoted therein.

As was emphasized Fundamental Theorem of asset Pricing describes arbitrage free markets phenomena by means of stochastic objects. In the finite-dimensional situation it is also well known that the corresponding description can be obtained by means of geometrical objects. The finite-dimensional geometric  criterium is given by Stiemke's Lemma (in particular, it implies Harrison's -- Pliska's theorem). The results of the article belong to this direction in a general Banach space situation (cf., in particular, Theorem~\ref{t.finite} and Remark~\ref{S-l1}).

 As a stimulating example we consider here a one-period financial market model   where arbitrage freeness criterium can be given in terms of  existence of a martingale measure which is equivalent to the initial one.
We show that the principal Banach space objects that possess   'arbitarge free' and 'martingale' geometric  behavior are plasterable cones and reflexive subspaces. Whereas  the main Banach geometry results constituting   mathematical foundation are Mazurs's convex sets separation theorem, Krasnosel'skij's description of plasterable cones and
Eberlein -- $\check{\text{S}}$mul'jan criterium for  reflexivity of a Banach space. Moreover it is uncovered that geometrically 'martingalness' can be expressed directly in terms of initial (not dual) objects and correspods to remoteness of a base of arbitrage possibilities from the financial strategies space (condition 2) of Theorem~\ref{t.osh-2}).

The article in fact forms a certain refinement and rearrangement of the material of  \cite{LebZab}. In addition we have  added a discussion of the results obtained in order to make their relation to other results and fields of analysis more transparent. The paper   is organized as follows. Preliminary Section~\ref{s.2} serves for introduction and explanation of geometrical objects that will be under analysis further. Here we recall a one-period market model and the corresponding Fundamental Theorem of asset Pricing  (Theorem~\ref{t.3}) and give  its geometric reformulation   (Theorem~\ref{t.4}) thus, in particular,  arriving at the equality $L\cap K = \{0\}$, where $L$ is a subspace and $K$ is a certain cone in a given Banach space $E$ as relation describing  absence of arbitrage; and relation $L^\perp \cap \tilde{K} \neq \emptyset$, where $\tilde{K}$ is a certain cone belonging to $K^*$ as describing  existence of a martingale measure. The goal of the paper is general Banach analysis  of objects of this sort and their interrelations.
An auxuliary Section~\ref{s.1'}
conains a collection and discussion of the known results on separation theorems, plasterable cones and reflexive subspaces that will be used in sequel.
 The main part of the article starts with  Section~\ref{s.5} which presents a Banach geometric picture of arbitrage absence   phenomena. Here
Theorem~\ref{t.osh-2} gives an alternative description of a martingale measure existence condition. It is proven that this condition can be given in a number of ways as in dual terms so also in direct geometric form (condition 2)\,) that does not contain dual objects (martingale measures).
It follows easily that  conditions obtained  are sufficient for arbitrage freeness (Corollary~\ref{t.osh-5}), but unfortunately, as is shown by example, in general they are not necessary conditions.
Complete Banach geometry  of arbitrage free markets (necessary and sufficient conditions) is given by  Theorem~\ref{t.osh-6} (the principal result of the article).  In particular, it shows that  assumption of finiteness of assets ($\dim L < \infty$) can be relaxed by means of   reflexivity condition ($L=L^{**}$) of the corresponding subspace.
These results describe the role of plasterable cones and reflexive subspaces in the whole of a play.
 As a by product on this base in a finite-dimensional situation we derive a certain additional information for Stiemke's Lemma (Theorem~\ref{t.finite} and Remark~\ref{S-l1}) and as
 an immediate corollary we also obtain a  refined version of the Fundamental Theorem of assets Pricing for the case considered (Theorem~\ref{t.4'}).
 We finish the article with  discussion in  Section~\ref{s.5'''1'} of
arbitrage freeness criteria for markets without any assumptions on    strategies subspace (i.e. no assumption on the nature of assets). Here the corresponding criteria can be obtained as in terms of initial objects (Theorem~\ref{t.9}) so also in dual terms (Theorem~\ref{t.bez-kr-dual}). However, contrary to Theorem~\ref{t.osh-6}, the latter criterium does not exploit martingale measures. It is formulated in the spirit of condition   4) of Theorem~\ref{t.osh-6} and in fact is related to the classical theorem on bipolar.

\section{Stimulating example -- one-period market model. Geometric formulation of The Fundamental Theorem of asset Pricing} \label{s.2}

This is a preliminary section. Its goal is to explain what sort of Banach structures will be under further analysis as objects  possessing   'arbitarge free' and 'martingale' geometric  behavior. Namely on the base of  a stimulating model example presented henceforth we show that
it is quite natural to consider equality $L\cap K = \{0\}$ \eqref{e.af}, where $L$ is a subspace and $K$ is a certain cone in a given Banach space $E$ as relation describing  absence of arbitrage; and relation $L^\perp \cap \tilde{K} \neq \emptyset$ \eqref{e.mm}, where $\tilde{K}$ is a certain cone belonging to $K^*$ as describing  existence of a martingale measure. The reader who takes this as self-evident (known) could immediately pass to the main part of the article which starts with  Section~\ref{s.5}.

As a stimulating example we  recall  the Fundamental Theorem of asset Pricing for one-period market model and give its Banach geometric reformulation (Theorem~\ref{t.4}).

A one-period  market model is described  in the following way. Let us denote by
\label{p.1}$\overline{\pi} := (\pi^0,\,\pi ):= (\pi^0,\pi^1, \ldots , \pi^d ) \in \mathbb{R}^{d+1}_+$ the (initial, known) price system at moment  $t_0$. By
$\overline{S}:= (S^0, \,S):=(S^0, S^1, \ldots , S^d)$ we denote the   price system at moment  $t_1$, that is a family of  {nonnegative} random variables on a probability space  $(\Omega, \cal{F}, P)$ \ (where $(\Omega, \cal{F}, P)$ is the space of  (possible) \emph{scenario}). It is assumed that all the random variables under consideration are summable, that is   $S^i \in L_1(\Omega, P), \ i=\overline{0,d};$ \, (in the corresponding field of Mathematical Finance it is normal to consider $L_0$ random variables, we assume here summability at least for two reasons: on the one hand in this way  we simply  incorporate Banach geometry in  our further analysis, and on the other hand this assumption in one-period model is quite natural -- without it one cannot even formulate the  Fundamental Theorem of asset Pricing for the situation under consideretion (Theorem~\ref{t.3})). The variable
   $S^0$ is assumed to be a riskless bond, that is  it is not random
\begin{equation}
 \label{e.0}
S^0:\equiv (1+r)\,\pi^0,
\end{equation}
  where $r$ is interpreted as a  bank interest rate (for purely mathematical reasons one can assume that  $r>-1$). In what follows we presuppose that
 \begin{equation}
 \label{e.1}
 \pi^0 =1,
 \end{equation}
that is the price  $\pi^0$ is normalized. Therefore
\begin{equation}
 \label{e.2}
S^0:\equiv (1+r);
\end{equation}
  A (starting) investment  \emph{portfolio} is a vector  \  $\overline{\xi}:= (\xi^0, \xi ):=(\xi^0, \xi^1, \ldots, \xi^d )\in \mathbb{R}^{d+1}$, where the values   $\xi^i$ can be negative.

\noindent The price of buying the portfolio (at moment $t_0$) is equal to
\begin{equation}
 \label{e.3}
\overline{\xi}\cdot\overline{\pi} := \sum_{i=0}^d\xi^i \, \pi^i\, .
\end{equation}
And the value of portfolio (at moment $t_1$) is the random variable
\begin{equation}
 \label{e.4}
\overline{\xi}\cdot\overline{S} = \sum_{i=0}^d\xi^i \, S^i (\omega) =  \xi^0 \, (1+r)+ {\xi}\cdot {S}\,.
\end{equation}
An \emph{arbitrage opportunity} is a portfolio  $\overline{\xi}\in \mathbb{R}^{d+1}$, such that
$$
\overline{\xi}\cdot\overline{\pi}\le 0, \ \ \text{but} \ \ \left\{ \overline{\xi}\cdot\overline{S} \ge_{\text{a.e.}}0 \ \ \text{and} \ \ P(\overline{\xi}\cdot\overline{S}> 0) \, >0\right\}.
$$
If the market is arbitrage free  (i.e. there are no portfolio satisfying the relation written above) it is reasonable to consider it as being just.

It is convenient to express the arbitrage freeness conditions in terms of the so-called  {discounted net gains}.
Recall that  \emph{discounted net gains} (at moment  $t_1$) are the random variables given by
\begin{equation}
\label{e.1'}
Y^i := \frac{S^i}{1+r} -\pi^i, \ \ \ i=1,\ldots,d\,.
\end{equation}
 Let us denote by  $Y$ the vector of discounted net gains
$Y:= (Y^1, \ldots, Y^d)$.

\noindent By  \eqref{e.0} we have   $Y^0 = \frac{S^0}{1+r} -\pi^0 =0$ and therefore  $Y^0$ does not play any role.
\begin{La}
\label{t.1}
{\rm [\cite{Fol-Sch}, Lemma~1.3 and condition (1.3)]} \ The following conditions are equivalent:

{\rm 1)} market is arbitrage free;

{\rm 2)} if  $\xi \in \mathbb{R}^d$ satisfies  $\xi\cdot Y \ge_{\text{a.e.}} 0$, then  $\xi\cdot Y =_{\text{a.e.}} 0$.
\end{La}
\smallskip

This lemma has clear geometric interpretation.

\noindent Let
\begin{equation}
\label{e.8}
 L := \{\xi\cdot Y, \ \xi \in  \mathbb{R}^d\} = \left\{ \sum_{i=1}^d \xi_i\, Y^i, \ \ (\xi_1, \ldots , \xi_d)\in \mathbb{R}^d\right\}
\end{equation}
be the subspace generated by the vectors  (functions) $Y^i, \ i=1,\ldots,d$. By  $L_{1+}$ we denote the cone of nonnegative functions
\begin{equation}
\label{e.0'}
L_{1+} := \{f\in L_1(\Omega, P), \ f\ge 0\}.
\end{equation}
 Lemma~\ref{t.1} means that
\begin{equation}
\label{e.5}
\text{\emph{a market is arbitrage free}} \ \Leftrightarrow \ \ L\cap L_{1+} = \{0\}.
\end{equation}
The foregoing observations make it natural to consider $L_{1+}$ as the cone of \emph{arbitrage possibilities}   (\emph{profit cone}) and consider $L$ as the subspace of \emph{financial market strategies}.
\smallskip

Evidently it is important to obtain description of (geometric, algebraic and etc.) conditions under which the equality
$L\cap L_{1+} = \{0\}$ takes place.  The most known market arbitrage freeness condition in financial mathematics is given below
in Theorem~\ref{t.3} (\emph{fundamental theorem of asset pricing}). To formulate this theorem we recall a number of notions.

A measure  $Q$ is said to be \emph{absolutely continuous} with respect to the initial measure  $P$ (the notation $Q\prec P$), if  $Q$ and $P$ are  defined on the same  $\sigma$-algebra  $\cal{F}$, and  $P(A) =0 \ \Rightarrow \ Q(A)=0$.

It is said that  $Q$ is \emph{equivalent} to  $P$ (the notation  $Q \approx P$), if  $Q\prec P$ and  $P\prec Q$ that is $P(A) =0 \ \Leftrightarrow \ Q(A)=0$.

By the Radon -- Nikodim theorem  we have that if  $Q\prec P$, then there exists a function  $\psi \in L_{1+}$, such that
$$
Q(A) = \int_A \psi (\omega) \, d\,P \ \text{for every}  \ A\in \cal{F}.
$$
This function $\psi$ is called the    \emph{Radon -- Nikodim  derivative} of measure   $Q$  with respect to measure $P$ and is denoted by  $\frac{d\,Q}{d\,P}$.
\begin{Tm}
\label{t.3}
{\rm [fundamental theorem of asset pricing; \cite{Fol-Sch}, Theorem~1.6]}

In terms of the objects described above the next relation takes place:
\smallskip

a market is arbitrage free  \ $\Leftrightarrow$ \ \ there exists a measure  $P^* \approx P$ with a bounded~$\frac{d\,P^*}{d\,P}$, such that
\begin{equation}
\label{e.6}
E_{P^*} (Y^i)= 0, \ i=1,\ldots, d\, ,
\end{equation}
here  $E_{P^*} (Y^i)$ is the expectation of  $Y_i$, i.e.
\begin{equation}
\label{e.06}
E_{P^*} (Y^i)= \int_\Omega Y_i \, d\,P^* = \int_\Omega Y_i \, \left(\frac{d\,P^*}{d\,P}\right) \,d\,P\,.
\end{equation}
\end{Tm}
\smallskip

If condition  \eqref{e.6} is satisfied then  $P^*$ is called a  \emph{martingale} (or \emph{risk-neutral}) measure.
\smallskip

Thus  Theorem~\ref{t.3} can be rewritten in the following way:
\smallskip

\noindent\emph{a market is arbitrage free  \ $\Leftrightarrow$ \ \ there exists a martingale measure  $P^* \approx P$ with a bounded~$\frac{d\,P^*}{d\,P}$.}

\medskip

Now let us implement a geometric reformulation of Theorem~\ref{t.3}. This will enable us to uncover in what follows  (in Section~\ref{s.1'}) a general geometric nature of this type phenomena and, in particular, to refine directly the mentioned fundamental theorem of asset pricing (see Theorem~\ref{t.4'}).
\smallskip

Let us consider a Banach space  $L_1(\Omega, P)$. As usually, elements of this space are equivalence classes of integrable functions, where the equivalence of two functions is given by their equality almost everywhere; and the norm is given by the integral.    Thus, all the equalities and inequalities are understood as 'almost everywhere'.

As is known, for the dual space  $L_1(\Omega, P)^*$  we have
$L_1(\Omega, P)^* = L_\infty (\Omega, P)$ (where $L_\infty (\Omega, P)$ is the Banach space of equivalence classes of essencially bounded functions with  essup--norm). In this case elements  $x^* \in L_\infty (\Omega, P)$ are identified with functionals  (elements of $L_1(\Omega, P)^*$) by means of coupling
\begin{equation}
\label{e.e61}
<x^*,u> = \int_\Omega u\, x^* \,{d\,P}, \ \ \ u\in L_1(\Omega, P).
\end{equation}
This relation also shows that a functional  $x^* \in L_1(\Omega, P)^*$ can be identified with an absolutely continuous with respect to  $P$  charge $F$, such that  $\frac{d\,F}{d\,P} = x^* \in L_\infty (\Omega, P)$, namely,
\begin{equation}
\label{e.e4}
<x^*,u> = \int_\Omega u\, {d\,F} = \int_\Omega u\, \left(\frac{d\,F}{d\,P}\right)\, {d\,P} =  \int_\Omega u\, x^* \,{d\,P}, \ \ \ u\in L_1(\Omega, P).
\end{equation}
On this base we identify
\begin{equation}
\label{e.62}
L_1(\Omega, P)^*\ni F \ \leftrightarrow \ \frac{d\,F}{d\,P} \in L_\infty (\Omega, P).
\end{equation}
Let us also note that condition  \eqref{e.6} (i.e. martingalness) is nothing else than the record
\begin{equation}
\label{e.7}
{P^*} \perp Y^i, \ \ i=1,\ldots, d\,,
\end{equation}
where  $Y^i\in L_1 (\Omega, P)$, \   $P^*$ is identified with the functional  $E_{P^*}$ on  $L_1 (\Omega, P)$ (see \eqref{e.06}), and  $\perp$ denotes orthogonality between  $P^*$ and $Y^i$, that is the equality  $<P^*,Y^i> =0$.

Clearly condition  \eqref{e.7} is equivalent to the condition
\begin{equation}
\label{e.7'}
{P^*} \perp L\,,
\end{equation}
 where $L$ is the subspace  \eqref{e.8} generated by the vectors   $Y^i, \ i=1,\ldots,d$; and \eqref{e.7'} is nothing else than the record
\begin{equation}
\label{e.7''}
{P^*} \in L^\perp,
\end{equation}
where $L^\perp \subset L_1(\Omega, P)^*$ is the subspace of functionals annihilating on  $L$.
\smallskip

Let us consider now the cone   $L_{1+}$  \eqref{e.0'} of nonnegative functions in  $L_1(\Omega, P)$. By  $L_{1+}^*\subset L_1(\Omega, P)^* = L_\infty (\Omega, P)$ we denote the cone of nonnegative functionals on $L_{1+}$, i.e.
\begin{equation}
\label{e.7'''}
L_{1+}^*:= \{{x^*} \in L_\infty (\Omega, P) : \, <x^*,u>\,  \ge 0 \  \text{for every} \ u\in L_{1+}\}.
\end{equation}
Evidently,  $L_{1+}^*$ coincides with the cone  $L_{\infty +}$ of nonnegative functions from  $L_\infty (\Omega, P)$, i.e.
\begin{equation}
\label{e.71'''}
L_{1+}^*= L_{\infty +}:= \{x^*\in  L_\infty (\Omega, P), \ x^*\ge 0\}.
\end{equation}
We denote by   $\tilde{L}_{\infty +}$ the cone
 \begin{equation}
\label{e.72'''}
\tilde{L}_{\infty +}:= \{x^*\in  L^\infty (\Omega, P), \ x^* >_{\text{a.e.}} 0\}.
\end{equation}
Recalling the identification of functionals with charges (cf. \eqref{e.e4}, \eqref{e.62}) we note that the
equivalence between  the initial measure  $P$  and a functional  $x^*\in L_\infty (\Omega, P)= L_1(\Omega, P)^*$ is recorded by the next relation
\begin{equation}
\label{e.73'''}
P\approx x^*   \     \Leftrightarrow  \ x^*\in  \tilde{L}_{\infty +}\,.
\end{equation}
\medskip

Now taking into account the record \eqref{e.7}, \eqref{e.7''}, \eqref{e.73'''} along with Lemma~\ref{t.1} (conditions \eqref{e.5}) one can rewrite Theorem~\ref{t.3} in the form of
\begin{Tm}
\label{t.4}{\rm [fundamental theorem of asset pricing: geometric formulation]}

 For the objects described above the following two conditions are equivalent:

{\rm 1)} $L\cap L_{1+} = \{0\}$ {\rm(}= absence of arbitrage{\rm)};

{\rm 2)} $L^\perp \cap \tilde{L}_{\infty +} \neq \emptyset$ {\rm(}= existence of a martingale measure{\rm)}.
\end{Tm}
The foregoing observations show that it is natural to consider equality
\begin{equation}
\label{e.af}
L\cap K = \{0\},
\end{equation}
 where $L$ is a subspace and $K$ is a certain cone in a given Banach space $E$ as relation describing  absence of arbitrage; and relation
 \begin{equation}
\label{e.mm}
L^\perp \cap \tilde{K} \neq \emptyset ,
\end{equation}
 where $\tilde{K}$ is a certain cone belonging to $K^*$ \, (${K}^*$ is the cone of nonnegative functionals on  $K$) as describing  existence of a martingale measure.

 It will be the starting point of our further analysis of  geometry of arbitrage free markets that will be implemented in the subsequent sections.
\smallskip

\section{Separation theorems, plasterable cones and reflexive subspaces}\label{s.1'}

To describe the geometric nature of arbitrage freeness, that is to find criteria for fulfilment of \eqref{e.af} we need a number of known results related to convex sets, separation theorems, cones and subspaces. For the sake of convenience of presentation we recall them in this auxuliary  section.
\smallskip

  Let $X$ be a topological linear space and  $A, \, B \subset X$. One says that a  linear continuous functional $l$ \emph{separates the sets} $A$ and $B$ if the following relation holds
\begin{equation}
\label{e.separ}
\inf_{u\in A} <l,u> \ \ge \ \sup_{u\in B} <l,u>.
\end{equation}
Here and henceforth  $<l,u>$ denotes the value of functional $l$ at point  $u$.

The basic result on convex sets separation is given by the next Mazur's theorem (see, for example, \cite{Dan-Sch}, Theorem~V.1.12)
 \begin{Tm}
\label{t.6} {\rm [Mazur]}
Let  $A$ and $B$ be convex nonintersecting sets in a topological linear space and $\overset{\circ}{A} \neq \emptyset$ \,{\rm(}$A$ has a nonempty interior{\rm)}. Then there exists a nonzero linear continuous functional separating these sets.
\end{Tm}

If  $B=L$ is a linear subspace then condition  \ $\sup_{u\in L} <l,u> \ <\infty$ \ is equivalent to condition  $<l,u> =0, \ u\in L$, i.e. $l\in L^\perp$. Under the satisfaction of this condition relation  \eqref{e.separ} transforms into inequality
$$
\inf_{u\in A} <l,u> \ \ge \ 0.
$$
Thus Mazur's theorem implies
\begin{Cy}
\label{cor.Mazur}
Let  $A$ be a convex set in a topological linear space $X$ such that   $\overset{\circ}{A} \neq \emptyset$
and  $L$ be a linear subspace of $X$. If $
A\cap L = \emptyset $, then there exists a nonzero continuous linear functional  $l\in L^\perp$, such that $
\inf_{u\in A} <l,u> \ \ge \ 0.
$
\end{Cy}
Note also an evident property: in Mazur's theorem  for every point  $u_0\in \overset{\circ}{A} $ and the separating functional  $l$ mentioned one has
$$
<l,u_0> \ > \ \sup_{u\in B} <l,u>;
$$
and so in Corollary~\ref{cor.Mazur} we have
\begin{equation}
\label{1.00}
<l,u_0> \ > \ 0.
\end{equation}

One more useful separation theorem which  in fact also follows from Mazur's theorem sounds as follows.
\begin{Tm}
\label{t.6'}
Let  $A$ and $B$ be convex closed nonintersecting sets in a topological linear space and  $B$ be a compact set.
Then there exists a nonzero linear continuous functional $l$ strongly separating these sets, i.e.
\begin{equation}
\label{e.separ'}
\inf_{u\in A} <l,u> \ > \ \sup_{u\in B} <l,u>.
\end{equation}
\end{Tm}
\medskip

 The initial objects in our further analysis are a Banach space   $E$ and a cone  $K\subset E$. Recall that  a \emph{cone} in a vector space is a set  $K$ possessing the following two properties:

1) $K$ is a convex set;

2) for every  $x\in K$ and any  $0<\lambda \in \mathbb R$ one has  $\lambda x\in K$.
\smallskip

 We denote by  ${K}^*$ the cone of nonnegative functionals on  $K$, i.e.
$$
{K}^*:= \{{x^*} \in E^* : \, <x^*,u>\,  \ge 0 \  \text{for any} \ u\in K\},
$$
here  $E^*$ is the space dual to  $E$.

 If  $K^*$ has a nonempty interior $\overset{\circ}{K^*} \neq \emptyset$ (that is $K^*$ is a \emph{solid} cone) then one can easily observe the next property:
 \begin{equation}
 \label{1.0}
\text{if} \ u\in E \ \text{is such that  for every} \ x^*\in \overset{\circ}{K^*} \ \text{one has} \ <x^*, u> \, \geq 0, \ \text{then} \ u\in \overline{K}.
 \end{equation}
Indeed. Firstly, it holds  $K^* = \left(\overline{K}\right)^*$, and in addition we have    $\overline{\overset{\circ}{K^*}} = K^*$. If  $u\notin \overline{K}$ then according to Theorem~\ref{t.6'} there exists a functional
$x^* \in \left(\overline{K}\right)^*=K^* $ such that  \ \mbox{$<x^*, u> \ < 0$}. Therefore for the functionals  $x^{*\prime} \in \overset{\circ}{K^*}$ that are sufficiently close to $x^*$ the inequality  $<x^{*\prime} , u> \ < 0$ holds as well.
\smallskip

Cones $K$ with the property
$\overset{\circ}{K^*} \neq \emptyset$ \, will play a principal role in the article
 and henceforth we recall the corresponding known results.
\smallskip

Let  $E$ be a Banach space and  $K\subset E$ be a certain nonzero cone.  The cone $K$ is called  \emph{plasterable} if there exist a cone $\tilde{K}$ such that $\tilde{K}\cap (-\tilde{K})= \{ 0\}$ and a positive number $\rho$ such that
$$
\text{for every} \ 0\neq x\in K \ \ B(x, \rho \Vert x\Vert)\subset \tilde{K},
$$
where $B(x, \delta):= \{y\in E : \Vert x-y\Vert < \delta\}$.
\smallskip

The next figure  illustrates the plasterability property.
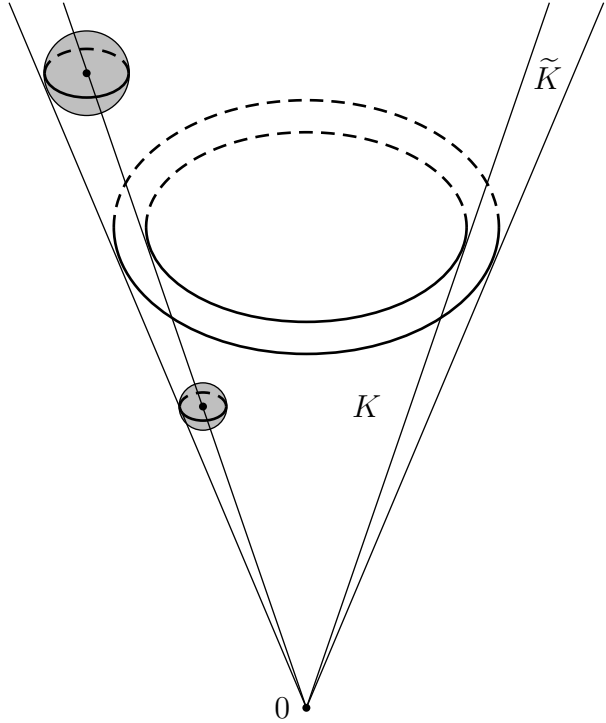
\psset{linewidth=0.5pt, unit=8.5mm,fillcolor=lightgray}
\begin{center}
	\begin{pspicture}(12.0,12.0)
	%\psgrid
	\psdot(6.0,0.5)
	\psellipticarc[linewidth=1.0pt,linestyle=dashed](6.0,8.0)(3.0,2.0){0}{180}
	\psellipticarc[linewidth=1.0pt](6.0,8.0)(3.0,2.0){180}{360}
	\psellipticarc[linewidth=1.0pt,linestyle=dashed](6.0,8.0)(2.5,1.5){0}{180}
	\psellipticarc[linewidth=1.0pt](6.0,8.0)(2.5,1.5){180}{360}
	\psline{-}(6.0,0.5)(10.6,11.5) \psline{-}(6.0,0.5)(9.76,11.5)
	\pscircle[fillstyle=solid](2.6,10.4){0.67}
	\psdot(2.6,10.4)
	\psellipticarc[linewidth=1.0pt,linestyle=dashed](2.6,10.4)(0.67,0.4){0}{180}
	\psellipticarc[linewidth=1.0pt](2.6,10.4)(0.67,0.4){180}{360}
	\pscircle[fillstyle=solid](4.4,5.2){0.38}
	
	\psellipticarc[linewidth=1.0pt,linestyle=dashed](4.4,5.2)(0.38,0.24){0}{180}
	\psellipticarc[linewidth=1.0pt](4.4,5.2)(0.38,0.24){180}{360}
	\psdot(4.4,5.2)
	\psline{-}(6.0,0.5)(1.4,11.5) \psline{-}(6.0,0.5)(2.24,11.5)
	\uput[0](5.3,0.5){$0$} \uput[0](6.5,5.2){$K$} \uput[0](9.3,10.4){$\widetilde{K}$}
	\end{pspicture}
\end{center}

A functional  $x^*\in E^*$ is said to be \emph{uniformly positive} on $K$ if there exists a constant  $c>0$, such that
\begin{equation}
\label{e.z001}
<x^*,u> \, \ge \, c\| u\|, \ \ u\in K.
\end{equation}

Let $F\subset E$ be a bounded, convex, and closed set that does not contain zero.  We denote by  $K(F)$  \emph{Krasnosel'skij's cone} that is the cone generated by vectors from  $F$. The set $F$ in this case is called
a  \emph{base} of $K(F)$.
\medskip

The next figure illustrates the notion of $K(F)$.
\psset{linewidth=0.5pt, unit=8.5mm,fillcolor=lightgray}
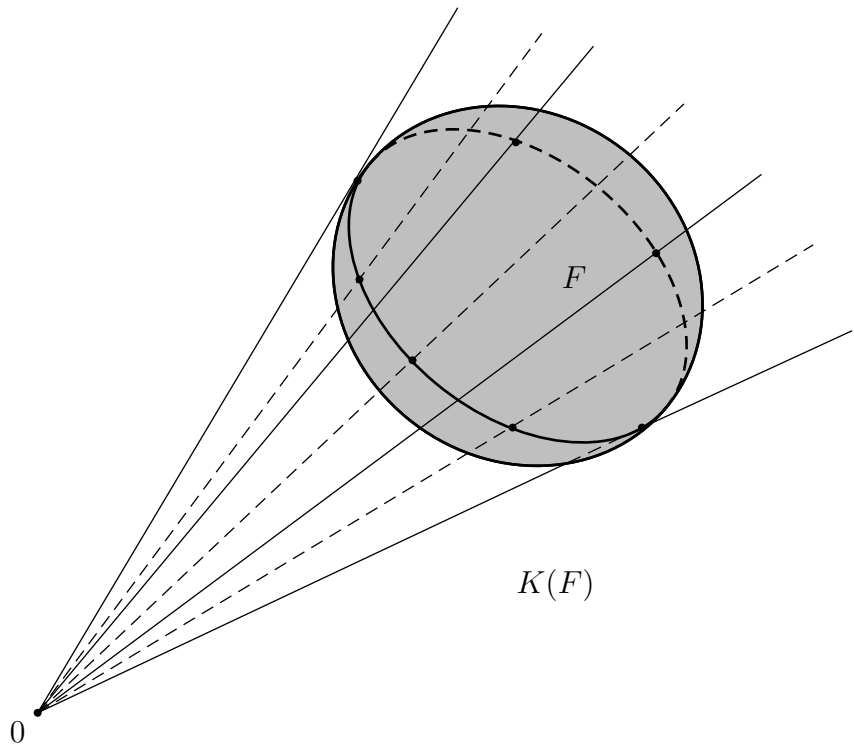
\begin{center}
	\begin{pspicture}(-2.0,0.0)(12.0,12.0)
	%\psgrid
	\psdot(-1.0,1.0)
	\rput[2.0,9.0]{-40}{\psellipse[linewidth=1.0pt,fillstyle=solid](0.0,10.0)(3.0,2.7)}
	\rput[2.0,9.0]{-40}{\psellipticarc[linewidth=1.0pt,linestyle=dashed](0.0,10.0)(3.0,2.0){0}{180}}
	\rput[2.0,9.0]{-40}{\psellipticarc[linewidth=1.0pt](0.0,10.0)(3.0,2.0){180}{360}}
	\rput[2.0,9.0]{-40}{\psellipse[linewidth=1.0pt](0.0,10.0)(3.0,2.7)}
	\psline{-}(-1.0,1.0)(5.5,12.0) \psline{-}(-1.0,1.0)(7.6,11.4)
	\psline{-}(-1.0,1.0)(10.2,9.4) %\psline{-}(-1.0,1.0)(11.6,6.96)
	\psdot(3.95,9.3) \psdot(6.4,9.9) \psdot(8.57,8.17) \psdot(8.35,5.45)
	\psline[linestyle=dashed]{-}(-1.0,1.0)(6.8,11.6) \psline[linestyle=dashed]{-}(-1.0,1.0)(9.0,10.5)
	\psline[linestyle=dashed]{-}(-1.0,1.0)(11.0,8.3)
	\psdot(3.98,7.76) \psdot(4.8,6.5) \psdot(6.35,5.45)
	\psline{-}(-1.0,1.0)(7.2,4.9)
	\psline[linestyle=dashed]{-}(7.2,4.9)(8.35,5.45)
	\psline{-}(8.35,5.45)(11.6,6.96)
	\uput[0](-1.64,0.7){$0$} \uput[0](6.9,7.8){$F$} \uput[0](6.2,3.0){$K(F)$}
	\end{pspicture}
\end{center}
Plasterable cones were introduced and studied in detail by Krasnosel'skij \cite{Kras1,KraLiSo}. Certain additional analysis is also implemented in  \cite{Bakh}.

The next theorem presents a number of criteria that characterize  property $\overset{\circ}{K^*}\neq \emptyset$ in terms of the objects introduced above (for the proof see \cite{Kras1} \S\S\ 4, 6; \cite{KraLiSo} \S 5 and \cite{Bakh}, Ch. 2, \S\S\ 8--10).
\begin{Tm}
	\label{t.oschtukatur'}
	Let $E$ be a Banach space and $K\subset E$ be a nonzero closed cone such that $K\cap (- K) = \{0\}$.
	The following six conditions are equivalent:
	
	{\rm 1)} \ $\overset{\circ}{K^*}\neq \emptyset$;
	
	{\rm 2)} \ $K$ is a plasterable cone;
	
	{\rm 3)} \ there exists a functional $x^*\in E^*$ which is uniformly positive on $K$ {\rm (}in fact  $\overset{\circ}{K^*}$  coincides with the set of such functionals{\rm)};
	
	{\rm 4)} \ there exists a convex bounded closed set $F$ not containing zero such that $K = K(F)$
	{\rm (}$K$ is Krasnoselskij's cone{\rm)};
	
	{\rm 5)} \ the set $F:=\overline{co}\left( K\cap \{u :\|u\|=1\}\right)$ \ {\rm(}where  $co (A)$ is the convex hull of  $A${\rm)} \ does not contain zero {\rm(}in this case $F$ is the base of $K${\rm)};
	
	{\rm 6)} \ if \, $x_n\in K, \ \|x_n\| = 1$, and $x_n$ converges weakly to $x_*$, then  $x_* \ne 0$.
\end{Tm}
\begin{Rk}\rm \ 1. Let  $K:=L_{p+}$ be the cone of nonnegative functions in $L_p, \ 1\le p \le \infty$. For $1\le p < \infty$ one has  $K^*=L_{p+}^* = L_{q+}$, where $\frac{1}{p} + \frac{1}{q} =1$ and for $q=\infty$ we assume that $\frac{1}{\infty} =0$.  Thus  $\overset{\circ}{K^*}=\overset{\circ}{L^*_{p+}}\neq \emptyset$ only when $p=1$.
	
	2. The foregoing remark  shows that property $\overset{\circ}{K^*} \neq \emptyset$ is rather special. On the other hand Theorem~\ref{t.oschtukatur'} tells  that in any Banach space there are a lot of cones possessing the mentioned property, namely one can take any
	Krasnosel'skij's cone (see condition 4).
\end{Rk}
\begin{Ee}
	\label{e.finite-plater}
	\rm Let $E= {\mathbb R}^n$. Note that a closed cone $K\subset {\mathbb R}^n$ is plasterable iff \
	\mbox{$K\cap (-K)= \{0\}$.}
	
	Indeed, let us verify that in  this situation condition 5) of Theorem~\ref{t.oschtukatur'} is satisfied, that is
	\begin{equation}
	\label{e.21}
	0\notin F, \ \ \ \text{where} \ \ F=\overline{co}\left( K\cap \{u :\|u\|=1\}\right)\,.
	\end{equation}
\end{Ee}
Recall that by Caratheodory's theorem on convex hull for every $y\in {co}\left( K\cap \{u :\|u\|=1\}\right)$ one has
\begin{equation}
\label{e.22}
y = \sum_{i=1}^{n+1}\lambda_i y_i\,,
\end{equation}
where $\lambda_i \ge 0, \ \  i\in \{1,\dots , n+1\}, \ \ \sum_{i=1}^{n+1}\lambda_i =1$, and $y_i \in \left( K\cap \{u :\|u\|=1\}\right)$.

Now if $x\in F$ then there exists a sequence $x_k \in {co}\left( K\cap \{u :\|u\|=1\}\right)$ such that
$$
x_k \to x.
$$
By \eqref{e.22} for every $x_k$ we have
\begin{equation}
\label{e.22}
x_k = \sum_{i=1}^{n+1}\gamma_{ik} x_{ik}\,,
\end{equation}
where $\gamma_{ik} \ge 0, \ \  i\in \{1,\dots , n+1\}, \ \ \sum_{i=1}^{n+1}\gamma_{ik} =1$, and $x_{ik} \in \left( K\cap \{u :\|u\|=1\}\right)$.

\noindent Since $\left(K\cap \{u :\|u\|=1\}\right)$ is a compact set we can assume (passing if necessary to a subsequence) that
\begin{equation}
\label{e.23}
\gamma_{ik} \to \gamma_i, \ \ \text{where} \ \ \gamma_i \ge 0, \ \  i\in \{1,\dots , n+1\}, \  \ \sum_{i=1}^{n+1}\gamma_{i} =1\,,
\end{equation}
and
\begin{equation}
\label{e.24}
x_{ik} \to x_i, \ \ \text{where} \ \  x_i \in \left(K\cap \{u :\|u\|=1\}\right), \ i\in \{1,\dots , n+1\}\,.
\end{equation}
That is any $x\in F$ has the form
\begin{equation}
\label{e.25}
x= \sum_{i=1}^{n+1}\gamma_{i} x_{i}\,.
\end{equation}
{where}  $\gamma_i \ge 0, \ \  i\in \{1,\dots , n+1\}, \   \sum_{i=1}^{n+1}\gamma_{i} =1$, \ and \
$x_i \in \left(K\cap \{u :\|u\|=1\}\right), \ i\in \{1,\dots , n+1\}$.

Now the proof goes by contradiction. Suppose that $x=0\in F$. If, for example, $\gamma_1 > 0$ in \eqref{e.25} then we have
$$
-K \ni - \gamma_1 x_{1} = \sum_{i=2}^{n+1}\gamma_{i} x_{i} \in K\,,
$$
that is
$$
0\neq - \gamma_1 x_{1} \in K\cap (-K)
$$
and we arrived at a contradiction with the assumption of example.
\smallskip

Among important objects in the article are also reflexive subspaces.
\smallskip

 Let  $L$ be a linear subspace of a Banach space $E$. We  call $L$ a \emph{reflexive subspace} and use the notation $L= L^{**}$ if for every linear continuous functional $h \in L^*$ (here $L^*$ is the dual space to $L$) there exists $x\in L$ such that
 \begin{equation}
 \label{1.0'}
h(f) = \ <f,x>,  \ \ f\in L^*.
 \end{equation}
By a  standard argument equality \eqref{1.0'} automatically implies the norms equality
\begin{equation}
 \label{1.0''}
\parallel h \parallel_{L^{**}} \ = \ \parallel x \parallel_L,
\end{equation}
that approves the term reflexive.
 \smallskip

  Note, in particular, that every finite dimensional subspace $L$ is reflexive.

A general description of reflexive subspaces is well known. It is based on Eberlein's -- $\check{\text{S}}$mul'jan's Banach space reflexivity     criterium.
\begin{Tm}
	\label{E-S}
	{\rm [Eberlein -- $\check{\text{S}}$mul'jan, \cite{Dan-Sch}, Theorem~V.4.7]} \ A Banach space is reflexive if and only if its closed unit ball is compact in weak topology.
\end{Tm}
This result in turn implies
\begin{Cy}
	\label{cor-E-S}
	A Banach space is reflexive if and only if any its bounded and weakly closed set is compact in weak topology.
\end{Cy}

\begin{Rk}
	\label{rk-1} {\rm If $E$ is a Banach space and $L$ is any its closed subspace then the closed unit ball of $L$ is nothing else than intersection of the closed unit ball of $E$ with $L$. Thus Theorem~\ref{E-S} implies the following  observations:
		
		1) \, a closed subspace $L\subset E$ is reflexive $L=L^{**}$ iff its closed unit ball is compact in weak topology;
		
		2) \, if $E$ is a reflexive Banach space then any its closed subspace $L$ is reflexive as well.}
\end{Rk}

Now we proceed to the main part  of the article.

\section{Arbitrage free markets geometry}\label{s.5}

We start with an alternative description of
 existence of a martingale measure condition.
\begin{Tm}
	\label{t.osh-2}
	Let  $E$ be a Banach space,  $K\subset E$ be a plasterable cone,  $F$  be any of its bases, and  $L\subset E$ be a linear subspace (not necessarily closed).
	The following three conditions are equivalent:
	
		{\rm 1)} $L^\perp \cap \overset{\circ}{K^*} \neq \emptyset$ \ \, {\rm(}= existence of a martingale measure 'refined condition'{\rm)};
	
	{\rm 2)} \ $\rho (F,L) > 0$, \, \ 	here  $\rho(A,B) = \inf_{x\in A, \, y\in B} \|x-y\|$;
	
	{\rm 3)}  $L^\perp + \overset{\circ}{K^*} = E^*$, \, here $E^*$ is the dual space to $E$.
\end{Tm}
\emph{Proof}. \ 1)\ $\Rightarrow$\ 2). Let $x^* \in L^\perp \cap \overset{\circ}{K^*}$. By Theorem~\ref{t.oschtukatur'} (condition 3)) \ $x^*$ is uniformly positive on $K$. Since $F$ is closed and $F\not\ni 0$ there exists $\delta >0$ such that  for every $u\in F$ one has $\Vert u \Vert >\delta$. Therefore for every $u\in F$ we have
$<x^*,u> \,\ge c\delta > 0$ where $c$ is the
constant mentioned in \eqref{e.z001}. Thus
$$
\rho (F,L)\ge \rho (\{u: <x^*,u> = c\delta \}, \{u: <x^*,u> = 0\}) > 0.
$$

2)\ $\Rightarrow$\ 1). Let $\rho (F,L) =\delta > 0$. Consider the set $M:=F + B\left(0, \frac{\delta}{2}\right)$. Clearly $M$ is a convex set, $M=\overset{\circ}{M}$, and $M\cap L =\emptyset$. Thus by Mazur's theorem (Theorem~\ref{t.6}) there exists a functional $x^*$ separating $M$ and $L$. One can easily verify that $x^* \in L^\perp \cap \overset{\circ}{K^*}$.
\smallskip

1) $\Rightarrow$ 3). \ Take any $x^* \in L^\perp \cap \overset{\circ}{K^*}$. For this $x^*$ there exists $\varepsilon >0$ such that
$$
B(x^*, \varepsilon) \subset \overset{\circ}{K^*},
$$
where $B(x^*, \varepsilon):= \{y^*\in E^* : \Vert x^*-y^*\Vert < \varepsilon\}$. Therefore
$$
L^\perp + \overset{\circ}{K^*}\supset (-x^*) + B(x^*, \varepsilon) = B(0, \varepsilon).
$$
And since $L^\perp + \overset{\circ}{K^*}$ is a cone it follows that $L^\perp + \overset{\circ}{K^*} = E^*$.
\smallskip

3) $\Rightarrow$ 1). \ The proof goes 'ad absurdum'.

Suppose that $ L^\perp \cap \overset{\circ}{K^*} = \emptyset$. Then by Mazur's theorem there exists $0\neq y^{**} \in (E^*)^*$ separating
$\overset{\circ}{K^*}$ and $L^\perp$. By the reasoning that follows Mazur's theorem one has
$$
y^{**} \in (L^\perp )^\perp \ \  \text{and} \ \ <y^{**},\overset{\circ}{K^*}>    \,
\geq 0,
$$
which in turn implies
$$
<y^{**},L^\perp +\overset{\circ}{K^*}> \, \geq 0
$$
and therefore $L^\perp + \overset{\circ}{K^*} \neq E^*$.

The proof is finished.
\medskip

It is natural to call  property  $\rho (F,L) > 0$  \emph{remoteness} of a base of arbitrage possibilities from the financial strategies space.
\smallskip

The term 'refined condition' used for the first condition of the theorem will be clarified in what follows (see Theorem~\ref{t.4'} and its comment).

As an immediate corollary of the theorem just proved one gets the next sufficient condition of market arbitrage freeness.

\begin{Cy}
	\label{t.osh-5}
	Let any of equivalent conditions 1), 2), or 3) of Theorem~\ref{t.osh-2} be fulfilled   then $L\cap K = \{0\}$  {\rm (}= the market is arbitrage free{\rm )}.
\end{Cy}
 Indeed.  Let, for example,  $x^* \in L^\perp \cap \overset{\circ}{K^*}$. Then $x^* (L) = 0$ and by condition 3) of Theorem~\ref{t.oschtukatur'}
we conclude that $x^* > 0$ on $K\setminus \{0\}$, that is $L\cap K = \{0\}$.
\smallskip

\begin{Rk}
	\label{r.1}
	\rm Unfortunately, as the next example shows,  for an arbitrary subspace $L$ the sufficient conditions just obtained are not necessary conditions.
\end{Rk}
{\bf Example}.  Let $E = l_1 = \{ (\xi_1, \xi_2, \dots ): \ \sum_i |\xi_i|<\infty\}$ and  let \ $K\subset l_1$ be the cone of nonnegative sequences and  $L$ be the subspace generated by  vectors of the form  \ $e_{2n}- \frac{1}{2n}e_{2n-1}, \ n=1,2, \dots$, \ where $e_k, k=1,2,\dots$ is the canonical base in  $l_1$.  Evidently $L$ is nothing else than the space of  vectors of the form  $\sum_{n=1}^\infty \xi_n (e_{2n}- \frac{1}{2n}e_{2n-1}):  \sum_{n=1}^\infty |\xi_n | <\infty$. Clearly
$$
K\cap L= \{0\}.
$$
The dual space to  $l_1$ is the space  $l_\infty$ of bounded sequences and the action of an element   $f = (\nu_1, \nu_2, \dots )\in l_\infty$ on $x= (\xi_1, \xi_2, \dots )\in l_1$ is given by the coupling
$f(x)= \sum_i \xi_i\,\nu_i$.
In this example   $K^*\subset l_\infty$ is the cone of nonnegative bounded sequences and  $\overset{\circ}{K^*}$ is the set of sequences separated from zero.
Moreover, if  $f = (\nu_1, \nu_2, \dots )\in L^\perp$ then    $\nu_{2n-1}= \frac{1}{2n}\nu_{2n}$.  This along with the boundness of the sequence  $(\nu_{2n})$ implies  $\nu_{2n-1}\to 0$. Therefore   $f\notin \overset{\circ}{K^*}$, that is
$$
L^\perp \cap \overset{\circ}{K^*} = \emptyset .
$$

The next (principal) result of the paper  distinguishes, in particular, the  class of subspaces for which the conditions mentioned in Theorem~\ref{t.osh-2} prove to be necessary  for market arbitrage freeness.

\begin{Tm}
	\label{t.osh-6}
	{\rm [arbitrage free markets geometry:
		plasterable profit cones   and reflexive subspaces]}
	Let $E$ be a Banach space,  $K\subset E$  be a plasterable cone,  $F$ be any of its bases, and  $L\subset E$ be a closed subspace such that its closed unit ball is compact in weak topology (i.e. $L$ is a reflexive subspace $L=L^{**}$). For the objects mentioned above the following four  conditions are equivalent:

	{\rm 1)} $L\cap K = \{0\}$ {\rm(}= absence of arbitrage{\rm)};
	
	{\rm 2)} \ $\rho (F,L) > 0$ {\rm (}= any base of arbitrage possibilities  is remote from   the market financial strategies space{\rm )};
	
	{\rm 3)}  $L^\perp \cap \overset{\circ}{K^*} \neq \emptyset$ {\rm(}= existence of a martingale measure 'refined condition'{\rm)};
	
	{\rm 4)}  $L^\perp + \overset{\circ}{K^*} = E^*$, where $E^*$ is the dual space to $E$.
\end{Tm}

\label{proof}
\emph{Proof}. \
\label{proof2}
By Theorem~\ref{t.osh-2} and Corollary~\ref{t.osh-5} it suffice to prove any of the implications 1)\ $\Rightarrow$\ 2), \ 1)\ $\Rightarrow$\ 3), \ or \ 1)\ $\Rightarrow$\ 4). Henceforce we give \emph{two} qualitatively different proofs of implication 1)\ $\Rightarrow$\ 3) each having (from our point of view) its own  value  and exploiting in different ways Mazur's theorem along with  Eberlein's -- $\check{\text{S}}$mul'jan's theorem,  Theorems~\ref{t.osh-2} and~\ref{t.oschtukatur'}.
\smallskip

\emph{The first proof of} \ 1)\ $\Rightarrow$\ 3).

Let us consider the set
\begin{equation}
\label{e.51'}
C:= \left\{ x^*|_L, \  x^*\in \overset{\circ}{K^*}\right\} \subset L^*,
\end{equation}
i.e. the set of restrictions of functionals from  $\overset{\circ}{K^*}$ onto  $L$.
\smallskip

\noindent Note that $C= \overset{\circ}{C}$ is an open set. Indeed. Let $C\ni h = x^*|_L$ where $x^* \in \overset{\circ}{K^*}$. By the Hahn -- Banach theorem  (on existence of norm nonincreasing extension of a functional) for every $g\in L^*$ with $\Vert g\Vert < \varepsilon$ there exists    $y^* \in E^*$ with $\Vert y^*\Vert < \varepsilon$ such that  $g =y^*|_L$. Now taking  sufficiently small $\varepsilon$  one has
$$
h + g = (x^* + y^*)|_L\, ,
$$
where $(x^* + y^*)\in \overset{\circ}{K^*}$. Thus $h + g \in C$.
\smallskip

\noindent Evidently,
\begin{equation}
\label{e.52'}
L^\perp \cap \overset{\circ}{K^*} \neq \{\emptyset\} \ \Leftrightarrow \ C \ni 0,
\end{equation}
where  $0\in L^*$.

\noindent Relations  \eqref{e.52'} show that to prove  1) $\Rightarrow$ 3) it is enough to verify relation
\begin{equation}
\label{e.53'}
0 \notin C \ \Rightarrow \  L \cap {K} \neq \{0\}.
\end{equation}
So henceforth we prove  \eqref{e.53'}.

\noindent Since  $\overset{\circ}{K^*}$ is a cone it follows that   $C=\overset{\circ}{C}$ is a cone in  $L^*$.

\noindent By assumption we have
$L=  L^{**}$.

\noindent As  $\overset{\circ}{C}$ is a cone and  $0 \notin \overset{\circ}{C} $ it follows (according to Mazur's theorem, along with equality  $L=  L^{**}$, and inequality  \eqref{1.00}) that there exists  $u_0 \in L$, such that
\begin{center}
	for every  $x^*\in \overset{\circ}{K^*} \ <x^*,u_0> \, > 0$.
\end{center}
On the one hand these inequalities mean that   $u_0 \in {K}$ (see  \eqref{1.0},  and recall that by assumption
we have  $K=\overline{K}$); and on the other hand they imply   $u_0 \neq 0$.
Thus,  1)\ $\Rightarrow$\ 3) is proved.
\smallskip

 \emph{The  second proof} of implication
 \ 1)\ $\Rightarrow$\ 3).
 \smallskip

 The proof goes by contradiction. Since $K$ is a plasterable cone one has  $\overset{\circ}{K^*}\neq \emptyset$  (see Theorem~\ref{t.oschtukatur'}). Suppose that  $L^\perp \cap \overset{\circ}{K^*} = \emptyset$.
 In this case, by Theorem~\ref{t.osh-2}, we have $\rho (F,L)=0$. This equality means that there exist sequences
 $u_n \in F$ and $v_n \in L$ such that
 \begin{equation}
 \label{e.31}
 u_n - v_n \to 0.
 \end{equation}
 Since $F$ is a bounded set the sequence $\{u_n\}$ is bounded as well and
 so in view of \eqref{e.31} the sequence  $\{v_n\}$ is also bounded.
 \smallskip

 By assumption $L$ is a reflexive subspace. Since $\{v_n\}$ is a bounded sequence one concludes by  Corollary~\ref{cor-E-S} (passing if necessary to a subsequence of $\{v_n\}$) that there exists a vector $v\in L$ such that $v_n \to v$ weakly, that is for every $x^* \in E^*$ we have
 \begin{equation}
 \label{e.31''}
 <x^*, v_n> \to v\,.
 \end{equation}
 This relation along with relation \eqref{e.31} implies that for every $x^* \in E^*$ we also have
 \begin{equation}
 \label{e.31'''}
 <x^*, u_n> \to v\,.
 \end{equation}
 Recall that $F$ is a closed convex set, so it is weakly closed (by Theorem~\ref{t.6'}). Therefore \eqref{e.31'''}
 means that $v\in F$. Note in addition that
 since $K$ is plasterable we have $0\not\in F$. So finally we have
 $$
 0\neq v \in (F\cap L) \subset (K\cap L),
 $$
 thus arriving at a contradiction. The proof is complete.
 \smallskip

Let us comment on conditions of Theorem~\ref{t.osh-6} and their relation to certain other fields of analysis.
\smallskip

Both conditions 3) and 4) of Theorem~\ref{t.osh-6} have deep and long mathematical history
(that does not only concern Mathematical Finance).
Namely, both of them are related to this or that sort of Separation Theorems (not only  Mazur's theorem, which serves as a corner stone of  the proof).

In particular, not to mention Mathematical Finance in connection with conditions 1) and 3) of Theorem~\ref{t.osh-6} may be it is reasonable to recall the classical Lagrange Multipliers Principle, which in Banach space terms can be  formulated as follows.
\begin{Tm}
\label{t.LMP} {\rm [Lagrange Multipliers Principle]}
	Let  $K$ be an open cone in a Banach space  $E$ and $L\subset E$ be a linear subspace. Then the following two conditions are equivalent.
	
	{\rm 1)} $L\cap K = \emptyset$,
	
	{\rm 2)} $L^\perp \cap K^* \neq \{ 0\}$.
\end{Tm}
  In optimisation problems  $K$ describes  \emph{increase directions} of a function that is optimized and  $L$ is the subspace of  \emph{tangent directions} of the domain at a point $x$ that is suspected for extremum. Thus here property 1) means that the function may attain minimum at $x$.  And property 2) is usually written in the form of  \emph{Lagrange Multipliers Equation}: there exist nonzero functionals  $\varphi_1\in L^\perp, \ \varphi_2\in K^*$ such that  $\varphi_1 +\varphi_2 =0$.
\smallskip

As for equivalence of conditions 1) and 4) of Theorem~\ref{t.osh-6} one could mention as a classical source theorem on bipolar (cf. Theorems~\ref{t.bez-kr-dual} and \ref{t.bez-kr-dual -refleks} of the present article) which in turn serves, in particular, as a mathematical instrument for the proof of the mentioned Lagrange Multipliers Principle.
\smallskip

Moreover, it should be emphasized that in a general situation  conditions of type  2), 3) and 4) of Theorem~\ref{t.osh-6}  are not equivalent: they have different mathematical nature: see, in particular, Section~\ref{s.5'''1'} of the article where condition 3) (martingalness) disapears at all while certain versions of conditions 2) and 4) are still in play, and therefore one can think of conditions 2) and  4) as of  'more stable' ones.

\begin{Rk}
	\label{r-2} {\rm Note that if $E$ is a reflexive space then by Remark~\ref{rk-1}
		one can completely relax condition on $L$ in the foregoing theorem, namely, in this case  $L$ can be any closed subspace of $E$.}
\end{Rk}
Continuing the preceding remark we observe in the next theorem that
one can relax  condition on $L$ not only  by means of the whole of the space $E$ but simply by   appropriate choice of  a profit cone $K$.
\begin{Tm}
	\label{t.osh-7}
	Let $E$ be a Banach space,  $K\subset E$  be a plasterable cone such that its base  $F$
	is compact in weak topology, and  $L\subset E$ be a closed subspace. For the objects mentioned above the following four  conditions are equivalent:

	{\rm 1)} $L\cap K = \{0\}$,
	
	{\rm 2)} \ $\rho (F,L) > 0$,
	
	{\rm 3)}  $L^\perp \cap \overset{\circ}{K^*} \neq \emptyset$,
	
	{\rm 4)}  $L^\perp + \overset{\circ}{K^*} = E^*$.
\end{Tm}
\emph{Proof}. By Theorem~\ref{t.osh-2} and Corollary~\ref{t.osh-5} one has \ 2)\ $\Leftrightarrow$\ 3)\ $\Leftrightarrow$\ 4)\ $\Rightarrow$\ 1).  So it is enough to verify, for example,  1)\ $\Rightarrow$\ 2). This moment is the only one where one needs weak compactness of the base $F$.

The proof goes by contradiction and is similar to  the second proof implication 1)\ $\Rightarrow$\ 3) given in the proof of Theorem~\ref{t.osh-6}.

  Suppose that $\rho (F,L) =0$. This means that
there exist sequences
$u_n \in F$ and $v_n \in L$ such that
\begin{equation}
	\label{e.31*}
	u_n - v_n \to 0.
\end{equation}
By weak compactness of  $F$ one concludes  (passing if necessary to a subsequence of $\{u_n\}$) that there exists a vector $u\in F$ such that $u_n \to u$ weakly, that is for every $x^* \in E^*$ we have
$$
<x^*, u_n> \to u\,.
$$
This relation along with relation \eqref{e.31*} implies that for every $x^* \in E^*$ we also have
$$
<x^*, v_n> \to u\,.
$$
$L$ being a closed subspace is weakly closed. Thus $u\in L$ and we obtain
$$
0\neq u \in (F\cap L) \subset (K\cap L)
$$
thus arriving to a contradiction.
\smallskip

 The foregoing results and discussion presented above
 convince us that Theorem~\ref{t.osh-6} manifests certain equilibrium between  conditions in the play. Namely, they show that once you start to shift one of these  conditions (either $L=L^{**}$ or $\overset{\circ}{K^*} \neq \emptyset$)
 the second one shifts as well (cf. Remark~\ref{r-2} and Theorem~\ref{t.osh-7})
 \smallskip

\smallskip
By  Example~\ref{e.finite-plater}  one has that in the case when $E={\mathbb R}^n$  a nonzero closed cone $K$ is plasterable iff $K\cap (-K)= \{0\}$.  This observation along with condition 5) of Theorem~\ref{t.oschtukatur'}
shows that in a finite-dimensional  situation Theorem~\ref{t.osh-6}
 looks as follows.
\begin{Tm}
	\label{t.finite}
		Let $ K\subset {\mathbb R}^n$ be a closed nonzero cone such that \ $K\cap (-K)= \{0\}$.  For a linear subspace  $L\subset {\mathbb R}^n$ the following four conditions are equivalent:

	{\rm 1)} $L\cap K = \{0\}$;
	
	{\rm 2)} $L\cap F = \{0\}$;
	
	{\rm 3)} $L^\perp \cap \overset{\circ}{K^*} \neq \emptyset$, where   $L^\perp$ the orthogonal
	complement to  $L$;
	
	{\rm 4)} $L^\perp + \overset{\circ}{K^*} = {\mathbb R}^n$.
\end{Tm}
\begin{Rk}
	\label{S-l1}
	\rm 1.  Condition 2) of Theorem~\ref{t.osh-6} turned into condition 2) of Theorem~\ref{t.finite} in view of compactness of the base $F$ in the situation under consideration.

	2.  Of course, here equivalence $1)\, \Leftrightarrow \, 3)$ is nothing else than Stiemke's Lemma for the situation considered.
\end{Rk}

Note also that as an immediate simple corollary of Theorem~\ref{t.osh-6} one can obtain a certain refinement of Theorem~\ref{t.4}.

Let us consider the cone
\begin{equation}
\label{e.72'''1}
\overset{\circ}{L}_{\infty +}= \{x^*\in  L_{\infty +},  \ x^* \ {\text{essentially separated from zero}} \ 0\}.
\end{equation}
Clearly,  $\overset{\circ}{L}_{\infty +}\subset \tilde{L}_{\infty +} $ \  and \  $\overset{\circ}{L}_{\infty +}$ is nothing else as the interior of the cone  $L_{\infty +}$ \eqref{e.71'''}.
\begin{Tm}
\label{t.4'}{\rm [refinement of Theorem~\ref{t.4}]}
Let $L\subset L_{1}$ be a closed linear reflexive subspace  $L=  L^{**}$ and $F:=\overline{co}\left( L_{1+}\cap \{u :\|u\|=1\}\right)$.
The following four conditions are equivalent:

{\rm 1)} $L\cap L_{1+} = \{0\}$ {\rm(}= absence of arbitrage{\rm)};

	{\rm 2)} \ $\rho (F,L) > 0$ {\rm (}= the base of arbitrage possibilities  is remote from   the market financial strategies space{\rm )};

{\rm 3)} $L^\perp \cap \overset{\circ}{L}_{\infty +} \neq \emptyset$ {\rm(}= existence of a martingale measure   {\rm(}refined condition{\rm)}{\rm)};

{\rm 4)}  $L^\perp + \overset{\circ}{L}_{\infty +}  = {L}_{\infty }$.
\end{Tm}

\emph{Proof}. \
 Take $E =L_1(\Omega, P)$ and $K=L_{1+}$. For this cone  we have $K^* = L_{1+}^*= L_{\infty +}$ (cf. \eqref{e.71'''}). Now apply Theorem~\ref{t.osh-6} keeping  in mind condition 5) of Theorem~\ref{t.oschtukatur'}.
\smallskip

\medskip

The figure illustrates the difference between condition 2) in Theorem~\ref{t.4} and condition 3) in Theorem~\ref{t.4'} (it also approves the term'refined condition').
\psset{linewidth=0.5pt, unit=8.5mm}
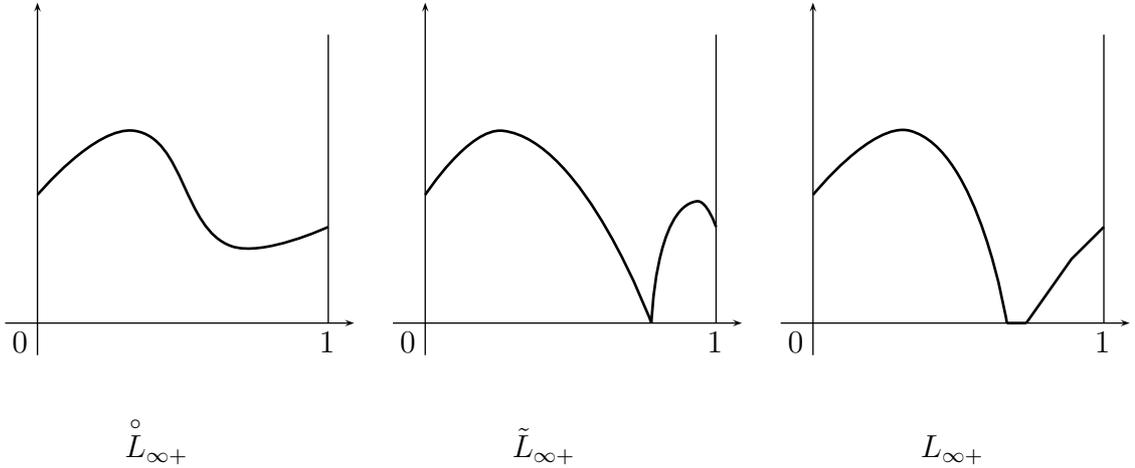
\begin{center}
\begin{pspicture}(18.0,6.0)
 %\psgrid
 \psline{->}(0.5,1.0)(5.9,1.0)  \psline{->}(1.0,0.5)(1.0,6.0)  \psline{-}(5.5,1.0)(5.5,5.5)
 \uput[0](0.4,0.7){$0$}  \uput[0](5.15,0.7){$1$}
 \pscurve[linewidth=1.0pt](1.0,3.0)(2.5,4.0)(4.0,2.2)(5.5,2.5)
 \psline{->}(6.5,1.0)(11.9,1.0)  \psline{->}(7.0,0.5)(7.0,6.0)  \psline{-}(11.5,1.0)(11.5,5.5)
 \uput[0](6.4,0.7){$0$}  \uput[0](11.15,0.7){$1$}
 \pscurve[linewidth=1.0pt](7.0,3.0)(8.2,4.0)(10.5,1.0)
 \pscurve[linewidth=1.0pt](10.5,1.0)(11.2,2.9)(11.5,2.5)
 \psline{->}(12.5,1.0)(17.9,1.0)  \psline{->}(13.0,0.5)(13.0,6.0)  \psline{-}(17.5,1.0)(17.5,5.5)
 \uput[0](12.4,0.7){$0$}  \uput[0](17.15,0.7){$1$}
 \pscurve[linewidth=1.0pt](13.0,3.0)(14.5,4.0)(16.0,1.0)
 \psline[linewidth=1.0pt](16.0,1.0)(16.3,1.0)
 \psline[linewidth=1.0pt](16.3,1.0)(17.0,2.0)(17.5,2.5)
\end{pspicture}
\end{center}
$$
\overset{\circ}{L}_{\infty +} \ \ \ \ \  \ \ \ \ \ \ \ \ \ \ \ \ \ \ \ \ \ \ \ \  \ \ \ \ \ \ \tilde{L}_{\infty +}\ \ \ \ \ \ \ \ \ \ \ \ \ \ \ \ \ \  \ \ \ \ \ \ \ \ \ \ \ \ \ \ \ L_{\infty  +}
$$
\medskip

Note also that  any finite-dimensional subspace $L$ is a reflexive subspace $L=  L^{**}$. Thus  Theorem~\ref{t.4} follows from the  equivalence of  1) \ $\Leftrightarrow$ \ 3)  in Theorem~\ref{t.4'}.

\begin{Rk}
\label{r1}
\rm Since  in  $l_1$  any closed infinite-dimensional subspace $L\subset l_1$ is not reflexive it follows that  in this case condition $L =L^{**}$ reduces to finite-dimensionality of $L$.  On the other hand   the situation with $L_1(\Omega, P)$ is qualitatively different: as is known even $\ell_2$ can be embedded in $L_1[0,1]$ (see in this connection, for example, \cite{AlbKal}, 5.6 and 6.4). Therefore, in general
condition $L =L^{**}$ of Theorem~\ref{t.4'} is essentially more general than condition $\dim L < \infty$ of Theorem~\ref{t.4}.
\end{Rk}

\section{Arbitrage freeness criteria for markets with arbitrary  financial market strategies subspace} \label{s.5'''1'}

This section forms a  certain 'philosophical' appendix of the paper. In the preceding section we obtained a number of market arbitrage freeness criteria under certain assumptions on the initial objects $K$ and $L$. Here we try to relax these assumptions. As we will see in  general situations the main role is played by analogies of  condition 4) of Theorem~\ref{t.osh-6} (dual objects;  cf. Theorem~\ref{t.bez-kr-dual}, \ref{t.bez-kr-dual -refleks}) and condition 2) of Theorem~\ref{t.osh-6} (initial objects; cf. Theorem~\ref{t.9}) while martingalness (condition 3) of Theorem~\ref{t.osh-6}) disappears. These observations also show, in particular, that in general conditions 2)   and 4) of Theorem~\ref{t.osh-6} are more 'stable'.  In addition the proof of Theorem~\ref{t.bez-kr-dual} explains tight relation between condition 4) of Theorem~\ref{t.osh-6} and classical theorem on bipolar.
\smallskip

We start with a description of market arbitrage freeness  by means of  dual objects in a situation when there are no  constraints on  $K$ and $L$ (thus here we do not presume any  assumption on the nature of profit cones and  assets).
\begin{Tm}
\label{t.bez-kr-dual}
{\rm [market arbitrage freeness criterium]} Let  $E$ be a Banach space and  $K,L \subset E$, where  $K$ is a closed cone, and $L$ is a closed subspace. The following two conditions are equivalent:

{\rm 1)} $L\cap K = \{0\}$ {\rm(}= absence of arbitrage{\rm)};

{\rm 2)}  \
$\text{\rm *-wcl}\, (K^* + L^\bot) = E^*$;

\noindent here  $E^*$ is the dual space to  $E$  and  $\text{\rm *-wcl}\, N$ denotes the closure of a set  $N$ in  *-weak topology of $E^*$.
\end{Tm}
\emph{Proof}. In fact this follows in a routine way from  theorem on bipolar (\cite{BourbakiTVS}, Proposition~IV.1.3.3). Indeed, as corollary of this theorem we have that for every family of closed cones $K_i, \ i=1, \dots , n$ \ one has
$$
\left(\bigcap_{i=1}^n K_i\right)^* = \text{\rm *-wcl}\,\left(\sum_{i=1}^n K_i^*\right),
$$
and therefore
$$
\left(L\cap K\right)^* =  \text{\rm *-wcl}\, (K^* + L^\bot)
$$
which finishes the proof.
\smallskip

Theorem~\ref{t.bez-kr-dual} formally gives an exhaustive answer on market arbitrage freeness condition in any situation. But of course in practice  verification of condition~2) is rather complicated. Let us only note here that as a corollary of Theorem~\ref{t.bez-kr-dual} one can obtain, in particular,  the next statement.
\begin{Tm}
\label{t.bez-kr-dual -refleks}
  Let  $E$ be a reflexive Banach space and  $K, L \subset E$, where  $K$ is a closed cone, and $L$ is a closed subspace. The following two conditions are equivalent:

{\rm 1)} \, $L \cap K = \{0\}$;

{\rm 2)} \, $\overline{K^*+L^\perp} = E^*$,

\noindent where in the latter condition 'over-line' means the norm closure in  $E^*$.
\end{Tm}

\emph{Proof}. \ For a reflexive space  $^*$-weak closure coincides with the weak closure. Moreover, since  $K^*+L^\perp$ is a  convex subset its weak closure coincides with the norm closure.
\smallskip

Theorem~\ref{t.bez-kr-dual} gives a description of absence of arbitrage condition in dual terms. However, one can
obtain a description of such markets without any usage of dual objects directly in the initial objects terms in the spirit of
 condition  2) of Theorem~\ref{t.osh-6}. This description is presented below in Theorem~\ref{t.9}, which in its turn is a corollary of the next result.

\begin{Tm}
\label{t.8}
Let  $E$ be a Banach space and  $K, L \subset E$, where  $K$ is a cone, and   $L$ is a linear subspace.

{\rm 1}. \ If for every  ${0\neq u\in K}$ one has  $(u +K) \cap L =\emptyset$, then  $L\cap K =\{0\}$.

{\rm 2}. \ Let  $K\cap (-K) = \{0\}$. In this situation
\smallskip

if  $L\cap K =\{0\}$, then for every  ${0\neq u\in K}$ one has  $(u +K) \cap L =\emptyset$.
\end{Tm}

We will say that  $K$ and $L$ are \emph{positively separated}, if for every  ${0\neq u\in K}$ one has  $(u +K) \cap L =\emptyset$.
\begin{Rk}
\label{r.3} \rm
Note that if condition 2) of Theorem~\ref{t.osh-6} is satisfied, that is $\rho (F,L) > 0$ then for every  ${0\neq u\in K}$ one also has $\rho (u +K,L) > 0$ (this can be verified, for example, by the argument similar to that exploited in the proof of  Theorem~\ref{t.osh-2}) and therefore separateness  of  $K$ and $L$ is weakening of the property of remoteness of $F$ from $L$.

\end{Rk}

\smallskip
Clearly Theorem~\ref{t.8} implies
\begin{Tm}
\label{t.9} {\rm [market arbitrage freeness criterium]} \
Let  $E$ be a Banach space and  $K, L \subset E$, where $K$ is a cone such that   $K\cap (-K) =\{0\}$, and   $L$ is a linear subspace.  The following two conditions are equivalent:

{\rm 1)} \ $L\cap K =\{0\}$ {\rm(}= absence of arbitrage{\rm)};

{\rm 2)} \ for every  ${0\neq u\in K}$ one has $(u +K) \cap L =\emptyset$ {\rm(}= $K$ and $L$ are positively separated{\rm)}.
\end{Tm}
\emph{Proof} of Theorem~\ref{t.8}. \  1. If  $0\neq u  \in L\cap K$, then  $L\ni 2u =u +u \in (u + K)$, that is  $(u +K) \cap L \neq\emptyset$.
\smallskip

2. The proof goes by contradiction. Suppose that there exists  ${0\neq u\in K}$, such that   $(u +K) \cap L \neq\emptyset$. It means that there exists a vector  $v\in K$, such that
$K\ni u + v = h \in L$. Since by assumption of theorem we have $L\cap K =\{0\}$, the latter relation means that $h=0$. This implies the equality
$K\ni v = - u \in (-K)$. So we arrived at a contradiction with the condition  $K\cap (-K) =\{0\}$.

\end{document}